\documentclass[aps,prl,twocolumn,showpacs]{revtex4}

\usepackage{graphicx}
\usepackage{dcolumn}
\usepackage{bm}
\usepackage[dvips]{epsfig}
\usepackage{natbib}
\usepackage{array}
\usepackage{inputenc}
\usepackage{xcolor}
\usepackage{multirow}
\usepackage{amsfonts}

\begin{document}

\title{Effective viscosity of microswimmer suspensions}
\author{Salima Rafa\"\i, Levan Jibuti, Philippe Peyla}
\affiliation{ Laboratoire de Spectrom\'etrie Physique, Grenoble,
France- UJF-CNRS UMR5588}

\begin{abstract}
The measurement of a quantitative and macroscopic parameter to
estimate the global motility of a large population of swimming
biological cells is a challenge Experiments on the rheology of
active suspensions have been performed. Effective viscosity of
sheared suspensions of live unicellular motile micro-algae
(\textit{Chlamydomonas Reinhardtii}) is far greater than for
suspensions containing the same volume fraction of dead cells and
suspensions show shear thinning behaviour. We relate these
macroscopic measurements to the orientation of individual swimming
cells under flow and discuss our results in the light of several
existing models.

\end{abstract}
%\begin{abstract}
%
%
%The measurement of a quantitative and macroscopic parameter to
%estimate the global motility of a large population of swimming
%biological cells is a challenge. In this letter, we show that the
%effective viscosity of sheared suspensions of billions of live
%unicellular motile micro-algae (\textit{Chlamydomonas Reinhardtii})
%is far greater than for suspensions containing the same volume
%fraction of dead cells. We also measure a shear thinning effect
%characteristic of complex fluids. We show that these two macroscopic
%observations are due to the microscopic behavior of motile live
%algae which are oriented during $70\%$ of the time parallel to the
%flow. This anisotropic orientation in the shear flow can be due to
%gravity or to a larger hydrodynamic aspect ratio of the chlamy due
%to flagella beating. We discuss our results versus several existing
%models. More generally, the experiments provide new insights into
%the rheology of suspensions of swimming micro-organisms.
%
%\end{abstract}

 \maketitle

 \begin{figure*}
 \begin{tabular}{lll}
 a & b & c \\
   \includegraphics[angle=0, width=0.5\columnwidth]{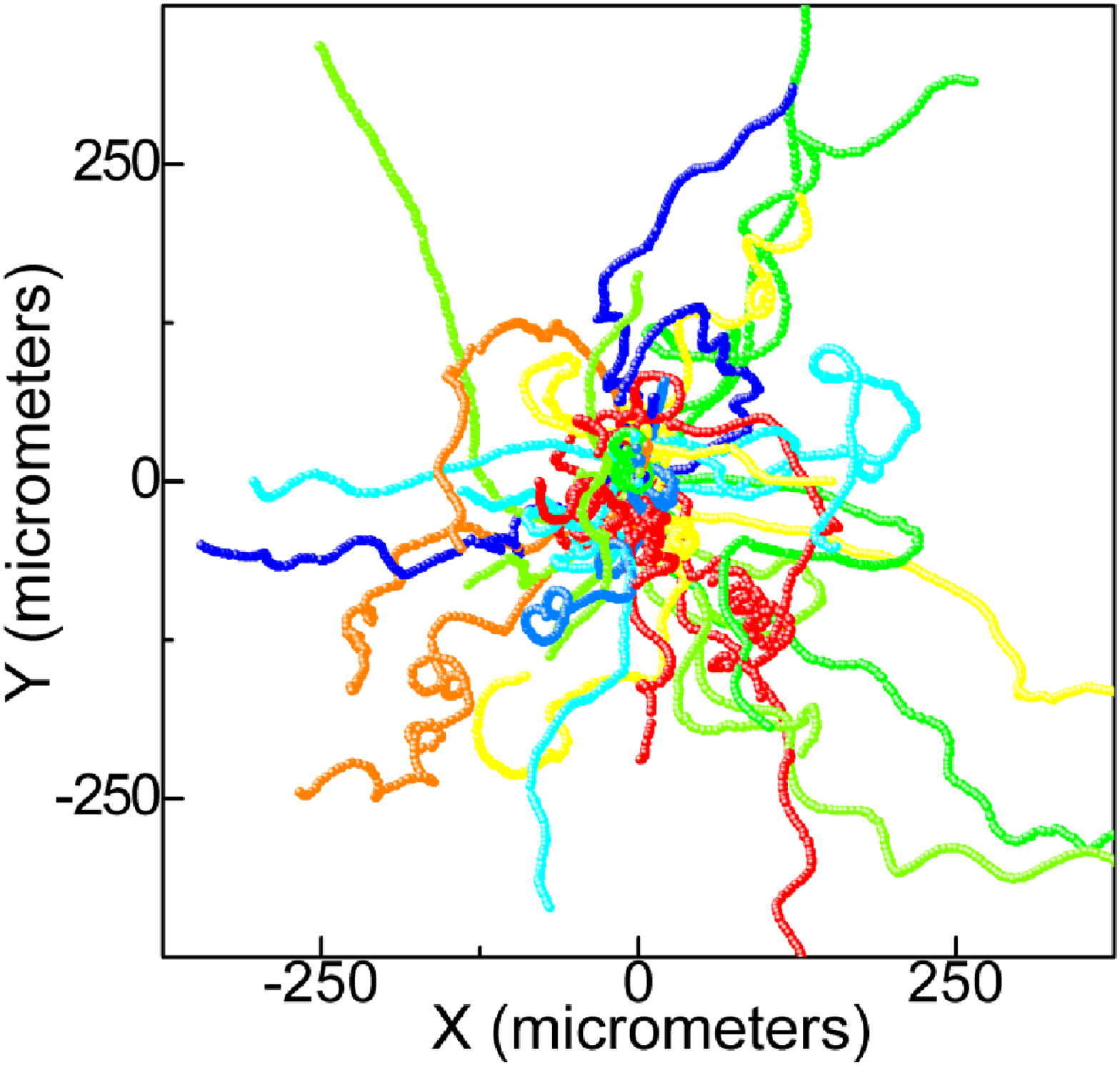} & \includegraphics[angle=0,width=0.63\columnwidth]{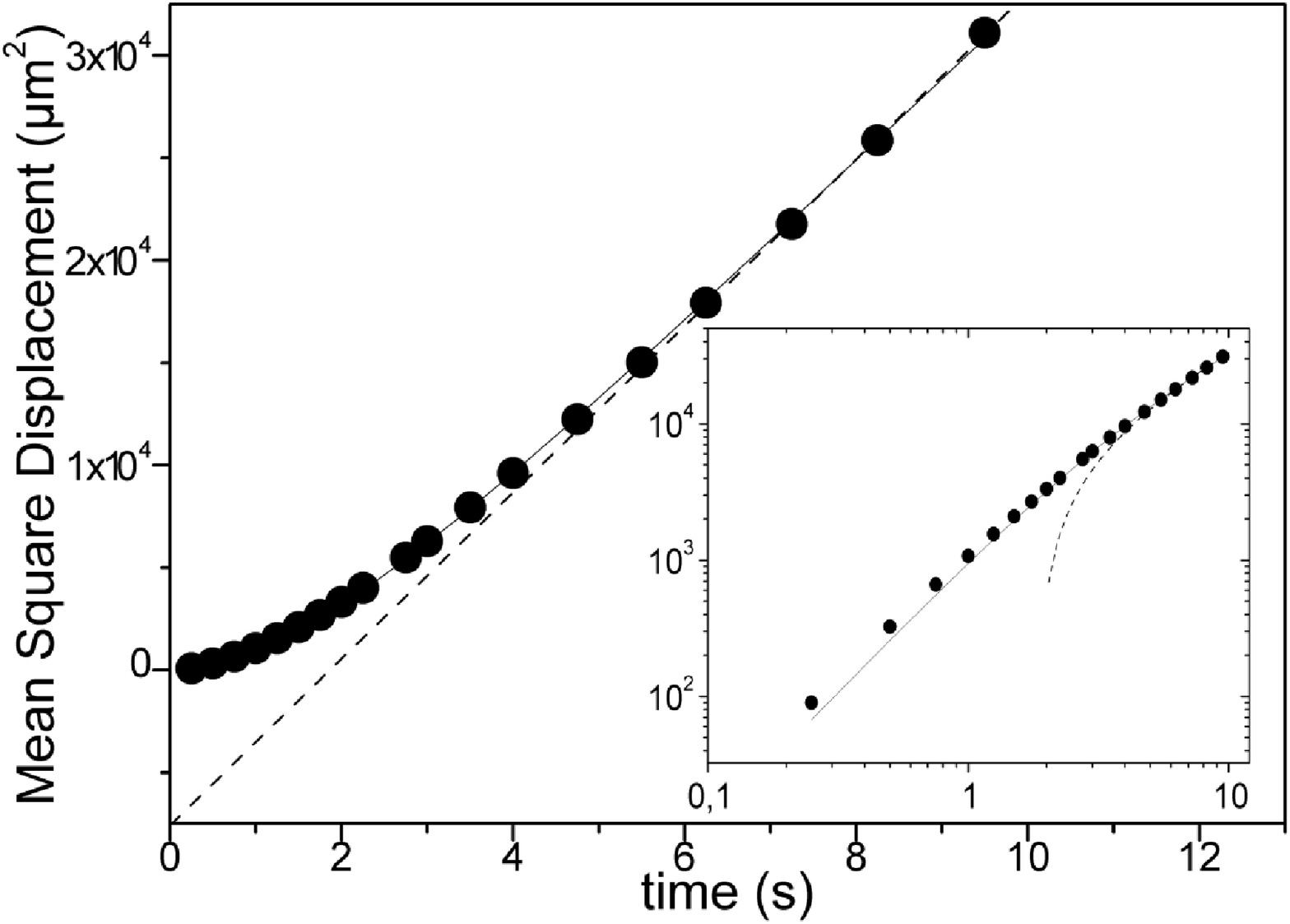} & \includegraphics[angle=0,
   width=0.63\columnwidth]{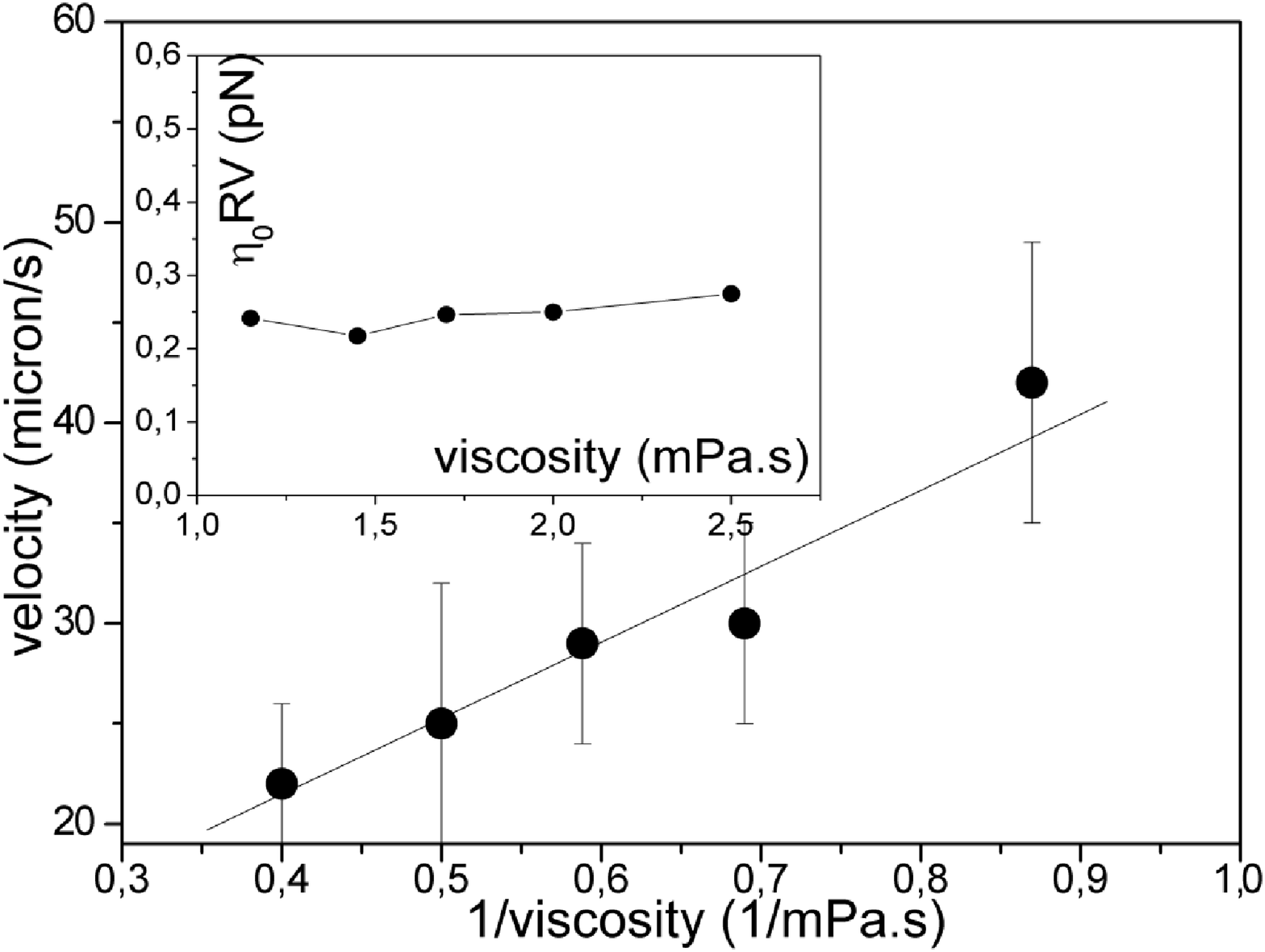}\\
 \end{tabular}
    \caption{a: Two dimensional trajectories of 50 swimming cells. Trajectory duration is 20 seconds. Trajectories starting positions have been all shifted to the origin.  b: Measured mean square displacement (MSD) of cells as a function of time. Averages have been extracted from 9Hz sequences of about 700 cells trajectories. MSD fits well to a persistent random walk process with a correlation time $t_{v}=3.5 \pm 0.1~s$ and a coefficient $D=995 \pm 20 ~\mu m^2.s^{-1}$(see text). Inset: Log-Log plot. c: Measured mean velocity $V$ of cells as a function of the inversed viscosity of the medium. The line represents a linear fit. Inset: Viscous drag force should be proportional to the product of the medium viscosity $\eta_0$, the mean radius $R$ and the velocity $V$ of the cells. This product $\eta_0 R V$ is shown to be independent of the medium viscosity. Error bars are smaller than the symbols. }
    \label{velovisco}
\end{figure*}

In nature, organisms that can propel themselves in a fluid medium
are ubiquitous. While larger organisms, such as fish, use inertia in
their motion, micro-organisms like spermatozoa, micro-algae or
bacteria, move at low Reynolds number, where viscous forces dominate
over the effects of inertia. A recent and currently unresolved issue
involves understanding the hydrodynamics associated with the
individual or collective motion of microswimmers through their
fluid-mediated interactions~\cite{Wulibchaber2000}. Microswimmer
suspensions have been shown to lead to complex dynamics such as the
so-called weak turbulence or bioconvection
phenomenon~\cite{Pedley1992, Mendelson1999, Dombrowski2004}. Such
active suspensions are made of self-propelled particles that create
a force multipole either at the front of the body, in which case
they are called pullers, or at the back, in which case, they are
pushers. The flow induced by the force multipole is responsible for
hydrodynamic interactions which are expected to have a dramatic
effect on dynamics and particularly on the rheology of the
suspension.

Recently, theoretical efforts have been made to model the effective
viscosity of active suspensions. Stokesian dynamic simulations by
Ishikawa and Pedley~\cite{Ishikawa2007} show a difference in
effective viscosity for suspensions of swimming spherical
`squirmers' in a gravity field. Haines et al.~\cite{Haines2008}
analytically showed that swimming results in a change in viscosity
if the orientation distribution of swimmers is assumed to be
anisotropic. More recently, Saintillan~\cite{Saintillan2009} used a
simple kinetic model to study the rheology of a dilute suspension of
self-propelled particles in a shear flow. He showed that suspensions
of pullers exhibit increased effective viscosity compared to passive
suspensions, while pusher suspensions exhibit a significant decrease
in viscosity due to the motile activity; these results are
consistent with previous predictions of  Hatwalne et
al.~\cite{ramaswamy} who used coarse-grained active hydrodynamics to
predict the rheology of active suspensions. Depending on swimmer
types (puller or pusher), cell shape (spherical or ellipsoidal) and
locomotion mechanisms, models can lead to very diverse results in
terms of rheology. In order to better understand the effects of
motility on viscosity, experimental work addressing the rheology of
microswimmer suspensions is necessary.

Very recently, Sokolov and Aranson~\cite{aranson2009}  measured the
microrheology of suspensions of pusher-like bacteria. They found, as
predicted by~\cite{Saintillan2009, ramaswamy}, that the effective
viscosity of such active suspensions decreases in comparison to
passive particles at the same volume fraction. To our knowledge, no
previous experimental data exist for puller-type micro-swimmer
suspensions for which an increase of effective viscosity is
predicted~\cite{Saintillan2009, ramaswamy}. This paper presents the
first direct experimental macroscopic measurement of the effective
viscosity of puller type microswimmer suspensions:
\textit{Chlamydomonas Reinhardtii} (CR), a 10 $\mu$m motile
unicellular alga. Rheological measurements performed on this system
show a clear increase in effective viscosity when compared to a dead
cell suspension. Shear thinning behaviour is also measured. Based on
models elaborated for dilute suspensions, we provide an
interpretation of our results, invoking the anisotropic distribution
of live cell orientations within the shear flow. We then discuss two
hypotheses that could account for the origin of this anisotropy: a
gravity torque or an effective elongated aspect ratio due to
flagella beating.

CR micro-algae~\cite{sourcebook} is a genus of green alga. It is a
bi-flagellated unicellular organism. Chlamydomonas is used as a
model organism for molecular biology, especially for studies of
flagella motion, chloroplast dynamics, biogenesis and genetics. They
are spheroidal in shape with two anterior flagella. Their
back-and-forth movement produces a jerky breast stroke with a mean
speed of $V \sim 40 \mu m/s$ in a water-like viscous medium. Since
the cell radius is $R\sim 5\mu m$, Brownian motion is negligible.
The swimming direction of the cells can be directed by stimulus
gradients: a phenomenon known as taxis, such as chemotaxis,
rheotaxis or phototaxis. Gradients are not used in our experiments
in order to avoid any external tropism on the motility. Wild-type
strains were obtained from the IBPC lab in Paris \cite{sandrine}.
Synchronous cultures of CR were grown in a Tris-Acetate Phosphate
medium (TAP) using a $14/10~hr$ light/dark cycle at $25^{o} C$.
Cultures were typically grown for two days under fluorescent
lighting before cells were harvested for experiments. These cultures
were concentrated up to a typical volume fraction of $20$ to $30\%$
by centrifuging for $20$ minutes at $900 g$ and re-suspending them
in a fresh culture medium to achieve the volume fractions required.
Volume fractions were measured using hematocrite capillaries and a
Neubauer counting chamber.

\begin{figure*}
 \begin{tabular}{ll}
 a & b\\
    \includegraphics[angle=0, width=0.62\columnwidth]{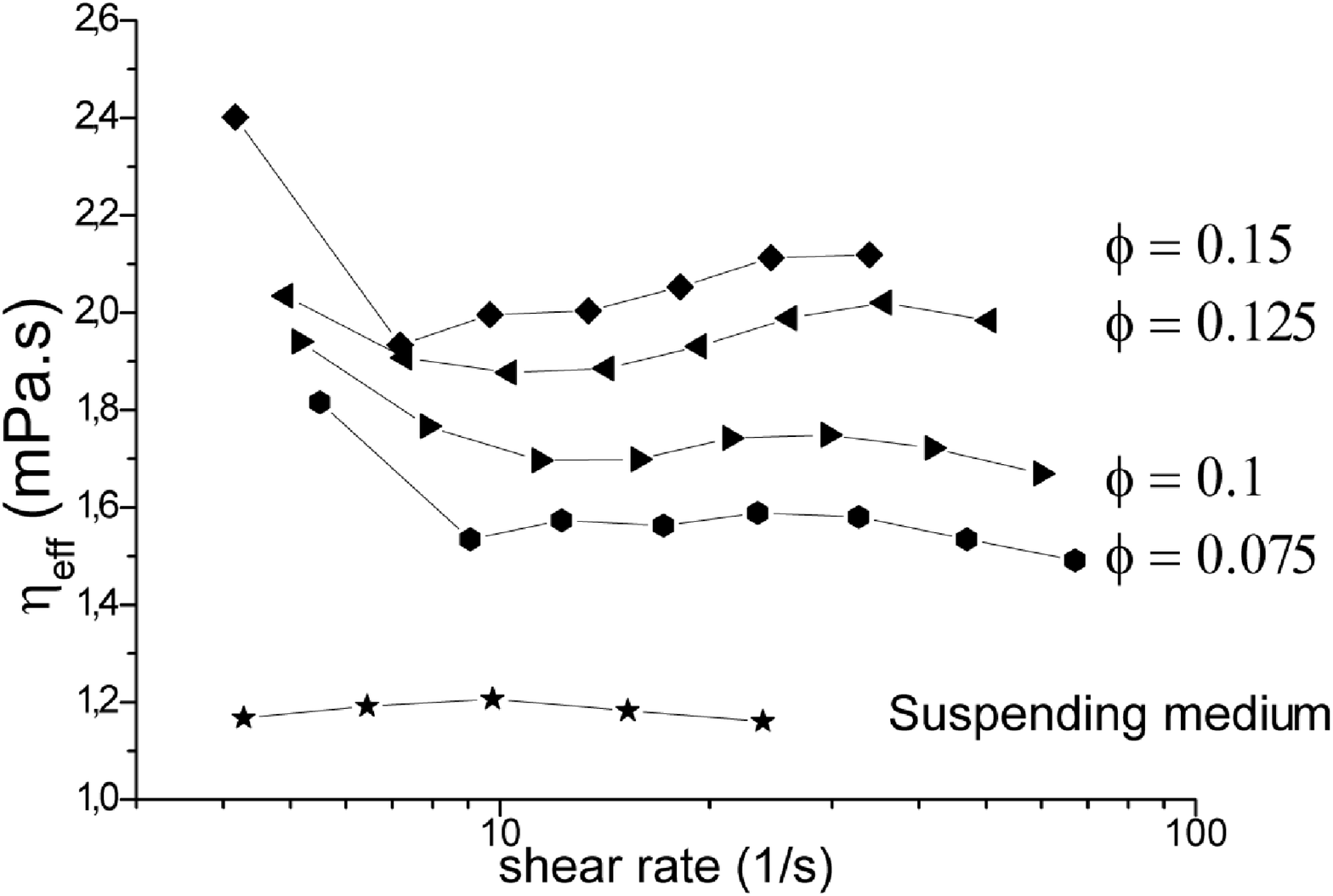}  & \includegraphics[angle=0, width=0.7\columnwidth]{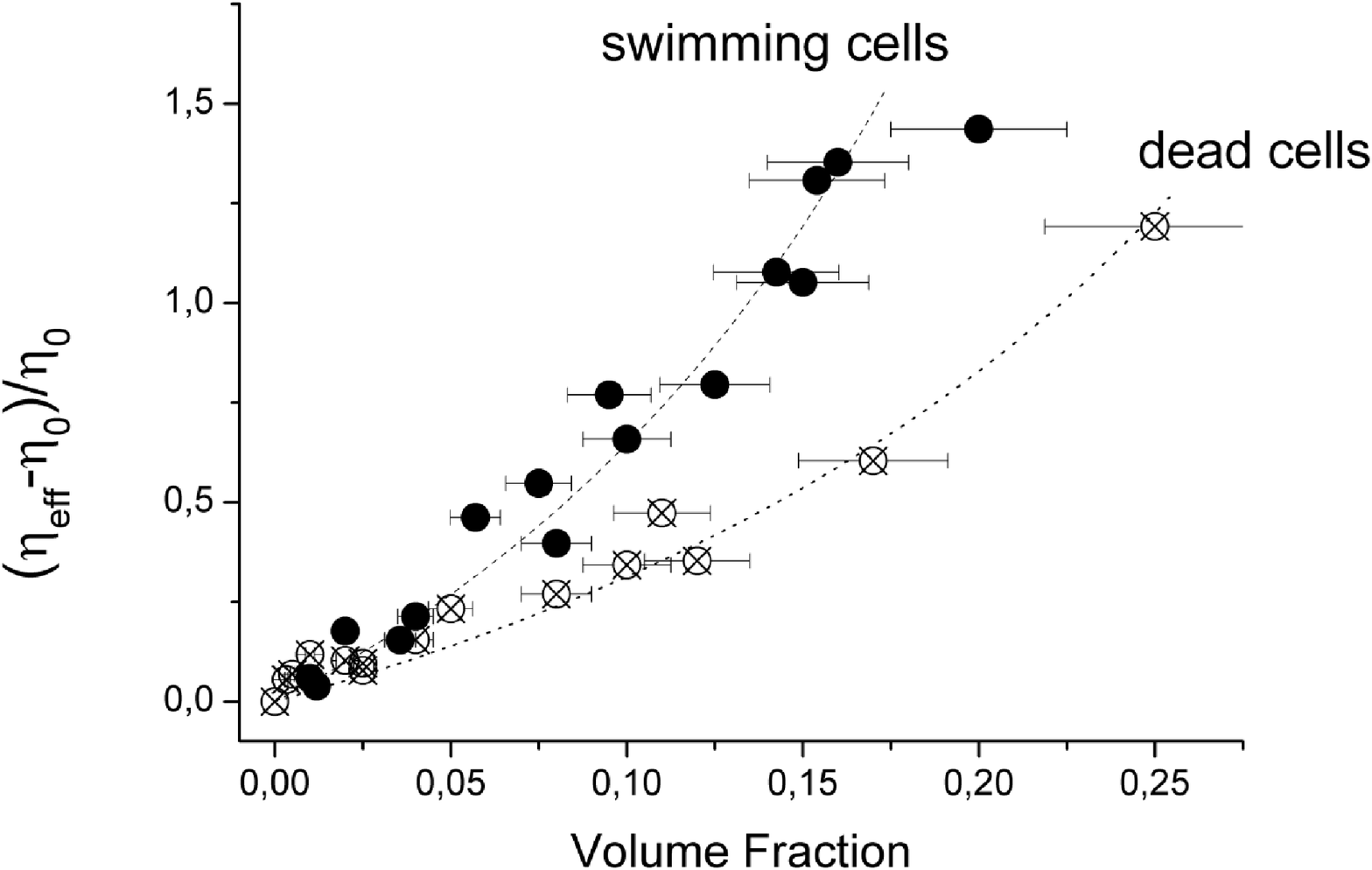}
\end{tabular}
    \caption{a. Effective viscosity of chlamydomonas suspensions as a function of shear rate. Data are shown for different volume fractions of the suspension and star symbols represent the viscosity $\eta_0$ of the culture medium. The lines are shown for ease of viewing only. b. Relative viscosity of microswimmer suspensions (measured at shear rate $=5 s^{-1}$) as a function of volume fraction. Solid symbols represent live cell data and crossed symbols represent dead cell data. Measurements are fitted to equation~\ref{eq:krieger} using $\phi_{\rm{m}}=0.62$ and $\alpha = $ 2.5 (dead cells) and 4.5 (swimming cells).}
    \label{rheo}
\end{figure*}

We started by characterizing cell motility. Microscopy imaging of
chlamydomonas suspensions was carried out on an Olympus inverted
microscope coupled with a Sensicam camera used at frame rates up to
30 Hz. Chambers made of glass and 240 $\mu m$ spacers were coated
with bovine serum albumine in order to reduce cell adhesion. Cells
were imaged in bright field using x10 magnification. Two dimension
trajectories of cells were recorded, typically for a few tens of
seconds. Particle tracking was performed using IDL~\cite{crocker}.
The cell trajectories observed are correctly modelled by a
persistent random walk~\cite{randommodels, Patlak1953}. Short term
correlations in the direction of movement are observed: the swimmer
then presents an almost fixed direction during a characteristic time
$t_{v}$. This short time stage corresponds to a ballistic regime
characterized by a velocity $V$. The ballistic regime ends when
swimmers make a turn. Note that the reorientation process has been
studied recently by R. Goldstein~\cite{PolinGoldstein2009} and is
due to a dephasing of the two flagella for a short time. Hence the
mean square displacement of cells is well described by a random walk
behaviour with a persistence length $\ell_p$ of approximately ten
cell diameters: $<x^2+y^2> = \ell_p^{2}~t/t_v -
0.5~\ell_p^{2}~[1-\exp{(-2t/t_{v})}]$ where $<>$ represents an
ensemble average over more than a thousand independent
measurements~\cite{crocker}. For $t<<t_v$, $<x^2+y^2> \sim V^2t^2$
with the ballistic velocity $V=\ell_p / t_v$ and for longer times a
random walk is observed, $<x^2+y^2> \sim 4Dt$ with $D=\ell_p^2 /4
t_v$.

The algae velocity dependency on viscous drag was measured by adding
a small amount of short chain dextran ($Mw=20000$) to the culture
medium. This allowed variation of the viscosity of the medium
between $1$ and $2.5 mPa.s$. We checked that swimming velocity is
inversely proportional to bath viscosity $\eta_0$
(fig.~\ref{velovisco}). This suggests that the stall force (the
force needed to stop the swimming cell), which is
proportional to the product $\eta_0 R V$, remains constant (fig.~\ref{velovisco}.c inset) %since the mean radius of the cells is independent of viscosity within the investigation range. This means that each alga swims by exerting a set of constant forces on the fluid.

Let us now turn to the actual viscosity measurements. The effective
viscosity of algae suspensions is measured on a Bohlin Gemini $150$
rheometer equipped with cone-plate geometry (cone angle = $2^o$,
diameter = $60$mm). Steady shear measurements were made at $T =
20^{o} C$. If a shear stress $\sigma$ is imposed and the associated
shear rate $\dot{\gamma}$ measured, effective viscosity is
$\eta_{\rm{eff}}=\sigma/\dot{\gamma}$. Samples of different volume
fractions were prepared and the effective viscosity measured for
both live and dead cells.  The cells were checked for motility after
the rheometric measurement.

The effective viscosity $\eta_{\rm{eff}}$ of active suspensions is
found to decrease with the shear rate (fig.~\ref{rheo}.a). As
discussed below, it reveals competition between the shear rate and
cells orientation in the flow. Note, that viscosity does not reach a
plateau value. This suggests that other effects like segregation
might play a role at higher shear rates. However, a clear global
decrease of $\eta_{eff}$ is observed from the maximum measured value
at $\dot{\gamma} \approx 4 s^{-1}$. We then investigated the
dependence of effective viscosity on the volume fraction of the
suspension. To do so, viscosity was measured at a given shear rate
of $5s^{-1}$, which is sufficiently high for rheometer resolution
but low enough to allow viscosity to be affected by motility.

Fig.~\ref{rheo}.b shows the relative viscosity
$(\eta_{\rm{eff}}-\eta_{0})/\eta_{0}$ of live and dead cell
suspensions as a function of the volume fraction. In both cases,
viscosity is an increasing function of the volume fraction as it is
for passive beads. Remarkably, the effective viscosity of swimming
cell suspensions is quantitatively larger than the viscosity of dead
cell suspensions (up to a factor of 2 for a 15\% volume fraction).

The effective viscosity $\eta_{\rm{eff}}$ of a suspension of passive
spherical particles in a solvent of viscosity $\eta_0$, depends on
its volume fraction $\phi$. Krieger and Dougherty's semi-empirical
law~\cite{Krieger1959} is shown to provide a reliable description of
the measurements:
\begin{equation}
\eta_{\rm{eff}} = \eta_0 (1-\frac{\phi}{\phi_m})^{-\alpha \phi_{m}}.
\label{eq:krieger}
\end{equation}

Here, $\phi_m$ is the maximal packing volume fraction. For a dilute
regime, where $\phi \ll 1$, eq.(\ref{eq:krieger}) reduces to
$\eta_{eff}\approx\eta_{0}(1+\alpha \phi)$, where $\alpha$ is known
as Einstein's intrinsic viscosity ($\alpha=2.5$ for passive and
spherical particles)~\cite{Einstein1906}.

To quantify the effect of motility on the effective viscosity of
cell suspensions, we fitted the measurements using
eq.(\ref{eq:krieger}). We set the maximal packing volume fraction to
$0.62$ and left $\alpha$ as a free parameter. This resulted in a
large difference between active and passive suspensions. We found
$\alpha =  2.5 \pm 0.1$ in the case of dead cells, which clearly
behave as passive particles. However, in the case of swimming cell
suspensions, we obtained $\alpha = 4.5\pm 0.2$. \emph{Swimming cells
induce a quantitative increase in the effective viscosity of the
suspension.}

\begin{figure*}[h]
\begin{tabular}{ll}
\includegraphics[angle=0,width=1.5\columnwidth]{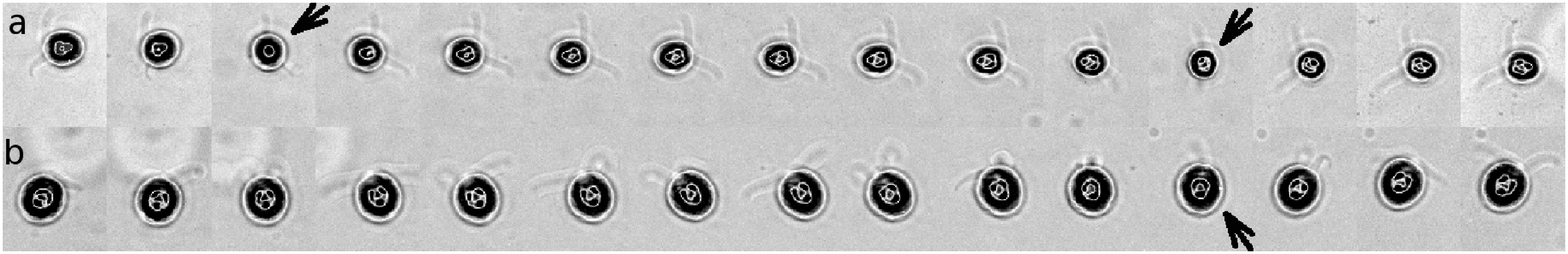}&\includegraphics[angle=0,
 width=0.4\columnwidth]{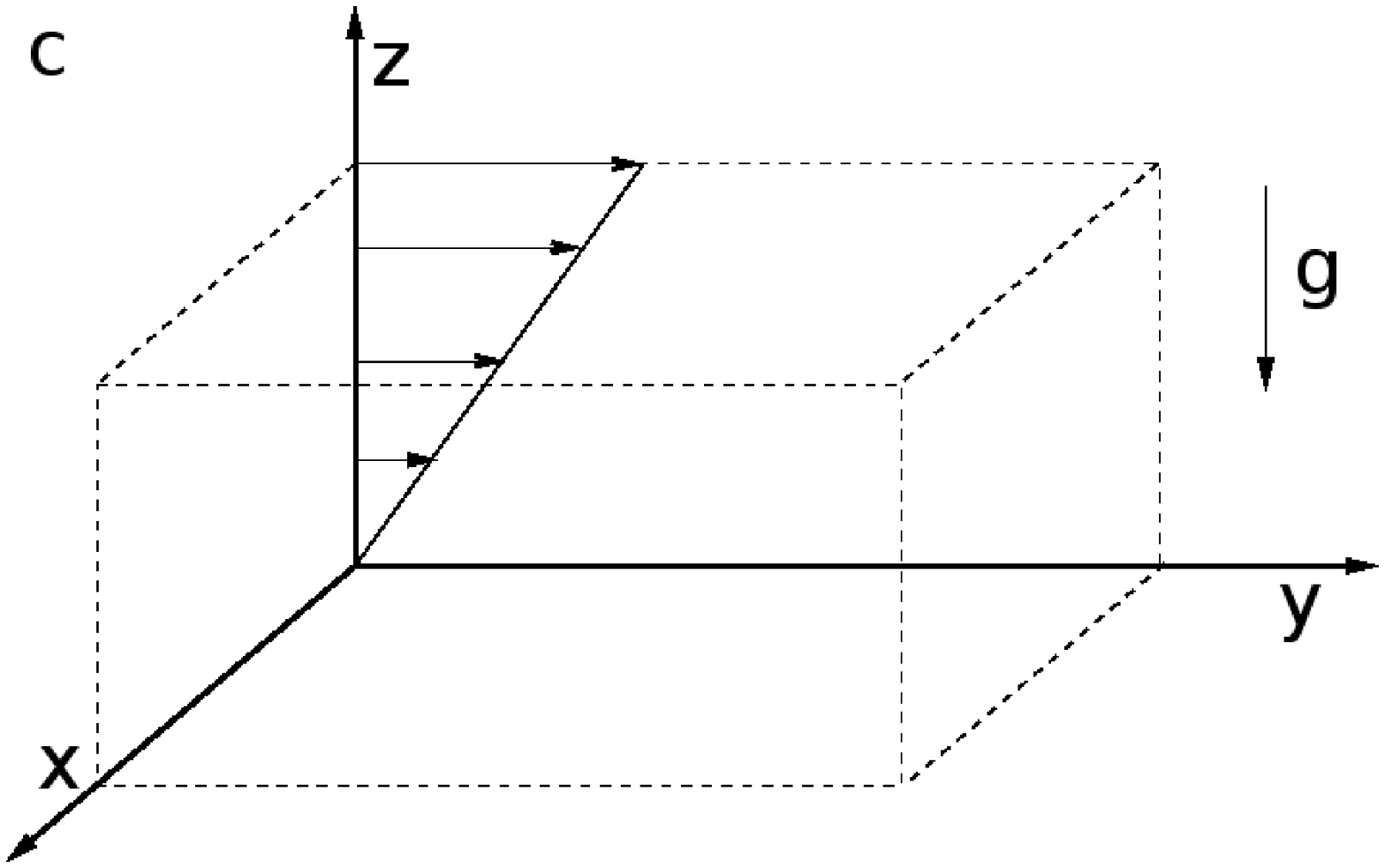}
\end{tabular}
    \caption{A. Dead cell in a 10 s$^{-1}$  shear flow experiencing tumbling around the $x-$axis.
      B. Live chlamydomonas swimming in the same shear flow. Arrows
      indicate cells flips and time between pictures is 20ms. The
      bar represents 10 micrometers.
   C. Schematic view of the flow cell, where $Oz$ is the direction of observation, $yOz$ is the shear plane.}
    \label{flow}
\end{figure*}

In order to understand this increase in effective viscosity, we
conducted complementary experiments which consisted of imaging the
cells while subjected to a shear flow. Dead cells show a regular
rotation at an angular velocity close to $\dot{\gamma}/2$, as would
passive spherical particles. We observed that swimming cells behave
very differently: they resist the flow rotation for most of the time
and eventually flip very rapidly. Fig.~\ref{flow}a and b show
picture sequences extracted from a fast-image film. High frequency
acquisition ($500 Hz$) allows us to determine whether or not an alga
is swimming by looking at the beating of flagella. In the case of a
dead cell, flagella only move because of thermal agitation, whereas
flagella of swimming cells beat at about $50 Hz$. Fig.~\ref{flow}
shows $50 Hz$ time sequences of cells subjected to a $10 s^{-1}$
shear rate. The swimming cell spends about two thirds of the period
in the $xOy-$plane and flips during one third of the period whereas
the dead cell rotates at a constant velocity close to
$\dot{\gamma}/2$.

Discussion. A puzzling question arises from our experiments: why
should live cells and dead cells with the same aspect ratio and the
same volume fraction respond so differently in a flow? In order to
answer such a question, we propose some hypotheses based on existing
models which might explain the increase of effective viscosity
according to individual dynamics of live cells. As shown on
fig.~\ref{velovisco}.b, the typical time $t_{v}$ after which a
chlamydomonas changes its direction of motion is about $3$ to $4$
seconds. This is much longer than the typical time associated with
the shear flow: $t_{f}=\dot{\gamma}^{-1}\approx 0.2 s$. This means
that live algae resist the flow rotation (fig. 3-A) during
approximately $70\%$ of the time $t_{v}$ by being aligned in the
flow. The fast flip between two aligned positions is indeed very
fast in comparison to the time the cell spends parallel to the flow.
After a time $t_{v}$, a reorientation occurs but the live alga is
then aligned in the flow once again and so on. On the contrary, a
dead cell rotates at a uniform angular velocity $\dot{\gamma}/2$
like a passive particle.

As proposed by Ramaswamy \cite{ramaswamy} and Saintillan
\cite{Saintillan2009}, when active particle distribution is
anisotropic (i.e. aligned in the flow) it gives rise to a modified
effective viscosity which is enhanced in the case of puller type
cells (and reduced in the case of pusher type cells
\cite{aranson2009}). Following~\cite{Saintillan2009, ramaswamy} a
swimming stress tensor can be defined:
$\mathbf{\Sigma}^{s}=\sigma_{0}[<\mathbf{p.p}>-\mathbb{I}/3]$.
$\sigma_0$ is the amplitude of the dipolar forces exerted by the
cell to swim (for pullers $\sigma_0>0$). The vector $\mathbf{p}$
represents the orientation of the cell and in our case, it is
oriented from the centre of each CR to the middle of the flagella
anchor points. Active suspensions with
$Q=<\mathbf{p.p}>-\mathbb{I}/3 \neq 0$ are thus equivalent to an
orientational ordered suspension with a non-zero deviatronic
swimming stress contribution. This contribution to the effective
shear viscosity is such that
$\eta_{eff}=\Sigma_{xy}^{s}/\dot{\gamma}$. Since for puller
$\sigma_{0}>0$, effective viscosity is increased. These models
although restricted to dilute regimes allow us to relate the
observed cell alignment in flow to the resulting increase in
viscosity.

We can think of two different hypotheses to explain the individual
dynamics observed (fig.\ref{flow}.b): either gravity effects or
large effective aspect ratio due to flagella beating. Gravity can
exert a torque on cells since the centre of mass of CR is known not
to coincide with its geometrical centre~\cite{Pedley1992}. Ishikawa
and Pedley~\cite{Ishikawa2007} showed that a suspension of
\emph{bottom heavy} squirmers results in an increased effective
viscosity. In 1970, Brenner~\cite{brenner1970} analytically
predicted the effect of gravity on the rheology of a dilute
suspension of inhomogeneous particles. He obtained for non rotating
particles an intrinsic viscosity $\alpha =\frac{5}{2}+\frac{3}{2}$.
This value is very close to our measurement of $\alpha$. Since dead
cells are observed to tumble like passive spherical particles, this
would suggest that mass distribution is homogeneous unlike live
cells. However, we have no experimental evidence that this is the
case. If the shear rate $\dot{\gamma}$ is increased, the constant
torque $T_G$ exerted by gravity will eventually become smaller than
the torque necessary to stop cell rotation $T_F = 4 \pi \eta R^{3}
\dot{\gamma}$. This would explain the shear thinning behaviour
observed and also predicted by Brenner~\cite{brenner1970}
in~fig.\ref{rheo}.a.

Another hypothesis that would explain the alignment of live cells is
that their systematic flagella beating results in an effective
hydrodynamic aspect ratio larger than that of their quasi-spherical
shape ~\cite{sriram}. The observed individual dynamics
(fig.\ref{flow}) would then correspond to Jeffery's orbits of
ellipsoidal prolate particles~\cite{jeffery1922} that result in an
increased effective viscosity. This is also predicted in
\cite{Saintillan2009, ramaswamy} where the puller suspension is
close to a suspension of passive Brownian rods for which it is well
known that a shear thinning behaviour can be observed
\cite{Saintillan2009} and \cite{hinchleal1976}. Note that Brownian
contribution is replaced here by hydrodynamic diffusivity due to
hydrodynamic interactions between swimming CR.

In this paper, we have experimentally shown that active suspensions
of puller type microswimmers present a dramatic increase in
effective viscosity. We have correlated these macroscopic
rheological measurements to the individual dynamics of cells under
shear flow: swimming cells tend to resist the flow rotation unlike
dead cells. More experiments are needed to explain the origin of
this phenomenon: a gravity torque exerted on the cell or an
effective large aspect ratio due to flagella beating.

 \begin{acknowledgments}
We thank C. Nizak for his help in the choice of the system and C.
Verdier for the rheometry. This work is financed by the Rhone-Alpes
Region through the Cible program.
 \end{acknowledgments}

% \bibliographystyle{apsrev}
%    %  \nocite{*}
%     \bibliography{chlamy}

\begin{thebibliography}{21}
\expandafter\ifx\csname
natexlab\endcsname\relax\def\natexlab#1{#1}\fi
\expandafter\ifx\csname bibnamefont\endcsname\relax
  \def\bibnamefont#1{#1}\fi
\expandafter\ifx\csname bibfnamefont\endcsname\relax
  \def\bibfnamefont#1{#1}\fi
\expandafter\ifx\csname citenamefont\endcsname\relax
  \def\citenamefont#1{#1}\fi
\expandafter\ifx\csname url\endcsname\relax
  \def\url#1{\texttt{#1}}\fi
\expandafter\ifx\csname urlprefix\endcsname\relax\def\urlprefix{URL
}\fi \providecommand{\bibinfo}[2]{#2}
\providecommand{\eprint}[2][]{\url{#2}}

\bibitem[{\citenamefont{Wu and Libchaber}(2000)}]{Wulibchaber2000}
\bibinfo{author}{\bibfnamefont{X.}~\bibnamefont{Wu}} \bibnamefont{and}
  \bibinfo{author}{\bibfnamefont{A.}~\bibnamefont{Libchaber}},
  \bibinfo{journal}{Physical Review Letters} \textbf{\bibinfo{volume}{84}},
  \bibinfo{pages}{3017} (\bibinfo{year}{2000}).

\bibitem[{\citenamefont{Pedley and Kessler}(1992)}]{Pedley1992}
\bibinfo{author}{\bibfnamefont{T.}~\bibnamefont{Pedley}} \bibnamefont{and}
  \bibinfo{author}{\bibfnamefont{J.}~\bibnamefont{Kessler}},
  \bibinfo{journal}{Annu. Rev. Fluid. Mech.} \textbf{\bibinfo{volume}{24}},
  \bibinfo{pages}{313} (\bibinfo{year}{1992}).

\bibitem[{\citenamefont{Mendelson et~al.}(1999)\citenamefont{Mendelson,
  Bourque, Wilkening, Andersoon, and Watkins}}]{Mendelson1999}
\bibinfo{author}{\bibfnamefont{N.}~\bibnamefont{Mendelson}},
  \bibinfo{author}{\bibfnamefont{A.}~\bibnamefont{Bourque}},
  \bibinfo{author}{\bibfnamefont{K.}~\bibnamefont{Wilkening}},
  \bibinfo{author}{\bibfnamefont{K.}~\bibnamefont{Andersoon}},
  \bibnamefont{and} \bibinfo{author}{\bibfnamefont{J.}~\bibnamefont{Watkins}},
  \bibinfo{journal}{J. Bacteriol.} \textbf{\bibinfo{volume}{181}},
  \bibinfo{pages}{600} (\bibinfo{year}{1999}).

\bibitem[{\citenamefont{Dombrowski et~al.}(2004)\citenamefont{Dombrowski,
  Cisneros, Chatkaew, Goldstein, and Kessler}}]{Dombrowski2004}
\bibinfo{author}{\bibfnamefont{C.}~\bibnamefont{Dombrowski}},
  \bibinfo{author}{\bibfnamefont{L.}~\bibnamefont{Cisneros}},
  \bibinfo{author}{\bibfnamefont{S.}~\bibnamefont{Chatkaew}},
  \bibinfo{author}{\bibfnamefont{R.~E.} \bibnamefont{Goldstein}},
  \bibnamefont{and} \bibinfo{author}{\bibfnamefont{J.~O.}
  \bibnamefont{Kessler}}, \bibinfo{journal}{Phys. Rev. Lett.}
  \textbf{\bibinfo{volume}{93}}, \bibinfo{pages}{098103}
  (\bibinfo{year}{2004}).

\bibitem[{\citenamefont{Ishikawa and Pedley}(2007)}]{Ishikawa2007}
\bibinfo{author}{\bibfnamefont{T.}~\bibnamefont{Ishikawa}} \bibnamefont{and}
  \bibinfo{author}{\bibfnamefont{T.~J.} \bibnamefont{Pedley}},
  \bibinfo{journal}{Journal of Fluid Mechanics} \textbf{\bibinfo{volume}{588}},
  \bibinfo{pages}{399} (\bibinfo{year}{2007}).

\bibitem[{\citenamefont{Haines et~al.}(2008)\citenamefont{Haines, Aranson,
  Berlyland, and Karpeev}}]{Haines2008}
\bibinfo{author}{\bibfnamefont{B.}~\bibnamefont{Haines}},
  \bibinfo{author}{\bibfnamefont{I.}~\bibnamefont{Aranson}},
  \bibinfo{author}{\bibfnamefont{L.}~\bibnamefont{Berlyland}},
  \bibnamefont{and} \bibinfo{author}{\bibfnamefont{D.}~\bibnamefont{Karpeev}},
  \bibinfo{journal}{Phys Biol} \textbf{\bibinfo{volume}{5}}, \bibinfo{pages}{9}
  (\bibinfo{year}{2008}).

\bibitem[{\citenamefont{Saintillan}(2009)}]{Saintillan2009}
\bibinfo{author}{\bibfnamefont{D.}~\bibnamefont{Saintillan}},
  \bibinfo{journal}{Exp. Mech} pp. \bibinfo{pages}{DOI
  10.1007/s11340--009--9267--0} (\bibinfo{year}{2009}).

\bibitem[{\citenamefont{Hatwalne et~al.}(2004)\citenamefont{Hatwalne,
  Ramaswamy, Rao, and Simha}}]{ramaswamy}
\bibinfo{author}{\bibfnamefont{Y.}~\bibnamefont{Hatwalne}},
  \bibinfo{author}{\bibfnamefont{S.}~\bibnamefont{Ramaswamy}},
  \bibinfo{author}{\bibfnamefont{M.}~\bibnamefont{Rao}}, \bibnamefont{and}
  \bibinfo{author}{\bibfnamefont{R.~A.} \bibnamefont{Simha}},
  \bibinfo{journal}{Phys. Rev. Lett.} \textbf{\bibinfo{volume}{92}},
  \bibinfo{pages}{118101} (\bibinfo{year}{2004}).

\bibitem[{\citenamefont{Sokolov and Aranson}(2009)}]{aranson2009}
\bibinfo{author}{\bibfnamefont{A.}~\bibnamefont{Sokolov}} \bibnamefont{and}
  \bibinfo{author}{\bibfnamefont{I.~S.} \bibnamefont{Aranson}},
  \bibinfo{journal}{Physical Review Letters} \textbf{\bibinfo{volume}{103}},
  \bibinfo{eid}{148101} (pages~\bibinfo{numpages}{4}) (\bibinfo{year}{2009}).

\bibitem[{\citenamefont{David~Stern}(2008)}]{sourcebook}
\bibinfo{editor}{\bibfnamefont{G.~W.} \bibnamefont{David~Stern},
  \bibfnamefont{Elizabeth~Harris}}, ed., \emph{\bibinfo{title}{The
  Chlamydomonas Sourcebook}} (\bibinfo{publisher}{Academic},
  \bibinfo{year}{2008}).

\bibitem[{san()}]{sandrine}
\bibinfo{note}{Physiologie Membranaire et Moléculaire du Chloroplaste, UMR
  7141, CNRS et Universit\'e Pierre et Marie Curie (Paris VI)}.

\bibitem[{\citenamefont{Crocker and Grier}(1996)}]{crocker}
\bibinfo{author}{\bibfnamefont{J.}~\bibnamefont{Crocker}} \bibnamefont{and}
  \bibinfo{author}{\bibfnamefont{D.}~\bibnamefont{Grier}},
  \bibinfo{journal}{Journal of Colloid and Interface Science}
  \textbf{\bibinfo{volume}{179}}, \bibinfo{pages}{298} (\bibinfo{year}{1996}).

\bibitem[{\citenamefont{Codling et~al.}(2008)\citenamefont{Codling, Plank, and
  Benhamou}}]{randommodels}
\bibinfo{author}{\bibfnamefont{E.}~\bibnamefont{Codling}},
  \bibinfo{author}{\bibfnamefont{M.}~\bibnamefont{Plank}}, \bibnamefont{and}
  \bibinfo{author}{\bibfnamefont{S.}~\bibnamefont{Benhamou}},
  \bibinfo{journal}{Journal of the Royal Society Interface}
  \textbf{\bibinfo{volume}{5}}, \bibinfo{pages}{813} (\bibinfo{year}{2008}).

\bibitem[{\citenamefont{Patlak}(1953)}]{Patlak1953}
\bibinfo{author}{\bibfnamefont{C.}~\bibnamefont{Patlak}},
  \bibinfo{journal}{Bulletin of Mathematical Biology}
  \textbf{\bibinfo{volume}{15}}, \bibinfo{pages}{311} (\bibinfo{year}{1953}).

\bibitem[{\citenamefont{Polin et~al.}(2009)\citenamefont{Polin, Tuval,
  Drescher, Gollub, and Goldstein}}]{PolinGoldstein2009}
\bibinfo{author}{\bibfnamefont{M.}~\bibnamefont{Polin}},
  \bibinfo{author}{\bibfnamefont{I.}~\bibnamefont{Tuval}},
  \bibinfo{author}{\bibfnamefont{K.}~\bibnamefont{Drescher}},
  \bibinfo{author}{\bibfnamefont{J.~P.} \bibnamefont{Gollub}},
  \bibnamefont{and} \bibinfo{author}{\bibfnamefont{R.~E.}
  \bibnamefont{Goldstein}}, \bibinfo{journal}{Science}
  \textbf{\bibinfo{volume}{325}}, \bibinfo{pages}{487} (\bibinfo{year}{2009}).

\bibitem[{\citenamefont{Krieger and Dougherty}(1959)}]{Krieger1959}
\bibinfo{author}{\bibfnamefont{I.~M.} \bibnamefont{Krieger}} \bibnamefont{and}
  \bibinfo{author}{\bibfnamefont{T.~J.} \bibnamefont{Dougherty}},
  \bibinfo{journal}{Trans. Soc. Rheol.} \textbf{\bibinfo{volume}{3}},
  \bibinfo{pages}{137} (\bibinfo{year}{1959}).

\bibitem[{\citenamefont{Einstein}(1906)}]{Einstein1906}
\bibinfo{author}{\bibfnamefont{A.}~\bibnamefont{Einstein}},
  \bibinfo{journal}{Ann. Phys.} \textbf{\bibinfo{volume}{19}},
  \bibinfo{pages}{289} (\bibinfo{year}{1906}).

\bibitem[{\citenamefont{Brenner}(1970)}]{brenner1970}
\bibinfo{author}{\bibfnamefont{H.}~\bibnamefont{Brenner}}, \bibinfo{journal}{J.
  of Colloid and Int. Science} \textbf{\bibinfo{volume}{32}},
  \bibinfo{pages}{141} (\bibinfo{year}{1970}).

\bibitem[{\citenamefont{Ramaswamy}()}]{sriram}
\bibinfo{author}{\bibfnamefont{S.}~\bibnamefont{Ramaswamy}},
  \bibinfo{note}{private communication: see
  http://www.condmatjournalclub.org/?p=760}.

\bibitem[{\citenamefont{Jeffery}(1922)}]{jeffery1922}
\bibinfo{author}{\bibfnamefont{G.}~\bibnamefont{Jeffery}},
  \bibinfo{journal}{Proceedings of the Royal Society of London}
  \textbf{\bibinfo{volume}{102}}, \bibinfo{pages}{161} (\bibinfo{year}{1922}).

\bibitem[{\citenamefont{Hinch and Leal}(1976)}]{hinchleal1976}
\bibinfo{author}{\bibfnamefont{E.}~\bibnamefont{Hinch}} \bibnamefont{and}
  \bibinfo{author}{\bibfnamefont{L.}~\bibnamefont{Leal}},
  \bibinfo{journal}{Journal of Fluid Mechanics} \textbf{\bibinfo{volume}{76}},
  \bibinfo{pages}{187} (\bibinfo{year}{1976}).

\end{thebibliography}

       \end{document}